\documentclass[11pt]{article}
\usepackage{verbatim,amsmath,amssymb,amsthm,graphicx,fullpage}

\newcommand{\markc}{\ensuremath{\mathsf{CW}}}
\newcommand{\markd}{\ensuremath{\mathsf{TP}_2}}
\newcommand{\marke}{\ensuremath{\mathsf{TP}_1}}
\newcommand{\items}{\ensuremath{\mathbb{I}}}
\newcommand{\types}{\ensuremath{\mathbb{T}}}

\newcommand{\opt}{\text{\small{\textsf{Opt}}}}

\newtheorem{proposition}{Proposition}
\newtheorem{corollary}[proposition]{Corollary}

\newtheorem{lemma}[proposition]{Lemma}
\newtheorem{theorem}{Theorem}

\theoremstyle{definition}
\newtheorem{definition}{Definition}
\newtheorem{example}{Example}

\DeclareMathOperator{\cost}{cost}
\newcommand{\ie}{\emph{i.e.}}
\newcommand{\eg}{\emph{e.g.},}
\newcommand{\etal}{\emph{et.~al.}}

\def\sfrac#1#2{\mbox{$\frac{#1}{#2}$}}

\begin{document}

\title{Online Companion Caching}

\author{Manor Mendel\\
Department of Computer Science\\ The Hebrew University, Jerusalem
\and
Steven S. Seiden%
\thanks{This research was partially supported by the Louisiana Board of 
Regents Research Competitiveness Subprogram and by AFOSR grant 
No. F49620-01-1-0264.}\\
Department of Computer Science\\ Louisiana State University, Baton Rouge}

\date{}

\maketitle

\begin{abstract}
This paper is concerned with online caching algorithms for the
$(n,k)$-companion cache, defined by Brehob {\etal}~\cite{BETW01}. In this
model the cache is composed of two components: a $k$-way set-associa\-tive
cache and a companion fully-associative cache of size $n$. We show that the
deterministic competitive ratio for this problem is $(n+1)(k+1)-1$, and the
randomized competitive ratio is $O(\log n \log k)$ and $\Omega(\log n +\log
k)$.
\end{abstract}

\begin{quote}
\emph{Steve Seiden died in a tragic accident on June 11, 2002. The first named
author would like to dedicate this paper to his memory.}
\end{quote}

\section{Introduction}

There is a rapidly growing disparity between computer processor speed
and computer memory speed. Of prime importance in bridging this gap is the \emph{cache},
the purpose of which is to allow quick access to memory items that are
accessed frequently. Since the cache is so important to system performance,
hardware designers have in recent years proposed a sequence of increasingly sophisticated cache designs (see \eg~\cite{J90,S93,CHKKSZ96}). Cache
designs can be conceptually thought as having two parts: An {\em architecture} and
a {\em caching algorithm}. The architecture describes the physical structure of
the cache such as its size and organization. The caching algorithm decides,
for a given sequence of requests for memory items, which items are
stored in the cache, and how they are organized, at each point in time.
While there is a large body of
theoretical work on caching algorithms for the simplest types of caches
(which we refer to as fully associative), little theoretical work has been
done on algorithms for more complicated cache architectures.
In this paper, we address this deficiency by providing the first 
theoretical analysis of the $(n,k)$-companion cache problem for $k>1$.

\paragraph*{Problem Description:}
A popular cache architecture is the \emph{set-associative cache}.
In a $k$-way set-associative cache, a cache of size $s$ is
divided into $m=s/k$ disjoint sets, each of size $k$. Addresses in
main memory are likewise assigned one of $m$ types, and the $i$'th
associative cache can only store memory cells whose address is
type $i$. Typically, there are $m=2^i$ such types, where the
$j$'th $k$-wise associative cache is indexed by $0\leq j \leq
2^i-1$ and memory addresses whose last $i$ bits are equal to $j$
are mapped to the $j$'th associative cache. Special cases includes
\emph{direct-mapped caches}, which are $1$-way set associative
caches, and fully-associative caches, which are $s$-way set
associative caches. Ideally $m$ should be small, but in order to
maintain the high speed of the cache, $k$ is usually very small.
1, 2 and 4-way caches are most commonly used.

\begin{figure}[htbp]
\centering
\fbox{
\includegraphics[width=4in]{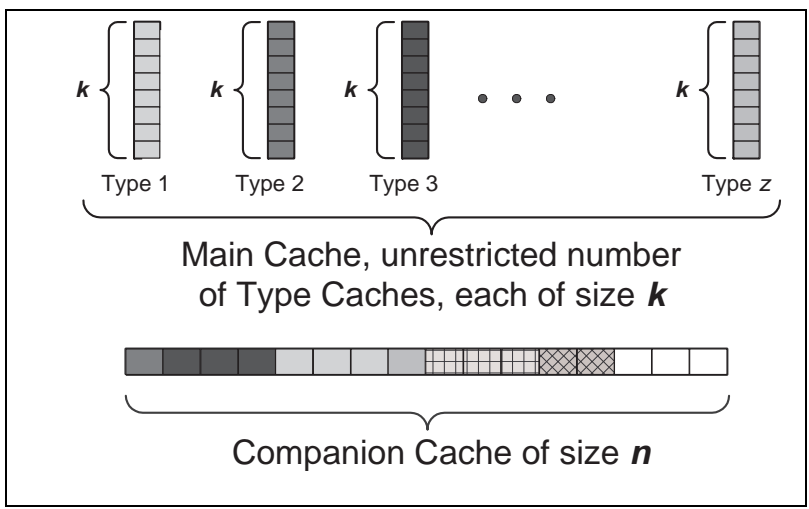}
}
\caption{A schematic description of a companion cache.}
\label{fig:cache}
\end{figure}

In order to overcome ``hot-spots", where the same set associative cache is
being constantly accessed, computer architects have designed hybrid cache
architectures. Typically such a cache has two or more components. A given
item can be placed in any of the components of the cache. Brehob
\etal~\cite{BWTE00,BETW01} considered the $(n,k)$ companion cache, which
consists of two components: A $k$-way set associative called the \emph{main
cache}, and a fully-associative cache of size $n$, called the
\emph{companion cache} (the names stem from the fact that typically $m k\gg
n$). As argued by Brehob \etal~\cite{BWTE00}, many of the L1-cache designs
suggested in recent years use companion caches as the underlying architecture.
Several variations on the basic companion cache
structure are possible. These include reorganization/no-reorganization and
bypassing/no-bypassing. Reorganization is the ability to move an item from
one cache component to another, whereas bypassing is the ability to avoid
storing an accessed item in the cache. A schematic view of the
companion cache is presented in Fig.~\ref{fig:cache}.

Since maintenance of the cache must be done online, and this makes it
impossible to service requests optimally, we use \emph{competitive
analysis}. The usual assumption is that any referenced item is brought into
the cache before it is accessed. Since items in the cache are accessed much
more quickly than those outside, we associate costs with servicing items as
follows: If the referenced item is already in the cache then we say that the
reference is a \emph{hit} and the cost is zero. Otherwise, we have a
\emph{fault} or \emph{miss} which costs one. Roughly speaking, an online
caching algorithm is called $r$-competitive if for any request sequence the
number of faults is at most $r$ times the number of faults of the optimal
offline algorithm, allowing a constant additive term.

\paragraph*{Previous Results:}
Maintenance of a fully associative cache of size $k$ is the well known
\emph{paging problem}~\cite{B66}. Sleator and Tarjan~\cite{ST85} proved that
natural algorithms such as \textsf{Least Recently Used} are $k$-competitive,
and that this is optimal for deterministic online algorithms. Fiat~{\etal}
\cite{FKLMSY91}, improved by McGeoch and Sleator~\cite{MS91} and
Achlioptas~{\etal}~\cite{ACN00}, show a tight $\approx \ln k$ competitive
randomized algorithm. $k$-way set associative caches can be viewed as a
collection of independent fully associative caches, each of size $k$, and
therefore they are uninteresting algorithmically.

Brehob {\etal}~\cite{BETW01} study deterministic online algorithms
for $(n,1)$-companion caches. They investigate the four previously
mentioned variants, i.e., bypassing/no-bypassing and
reorganization/no-reorganization.

\begin{table}[ht]
\begin{center}
\begin{tabular}{|ccccc|}
\multicolumn{5}{l}{Previous Results \cite{BETW01} (only for main cache of size $k=1$):}  \\
\hline   Bypass & Reorg' & det'/rand'  & Upper Bound & Lower Bound   \\ \hline
  $-$  & $-$ & det & $2n+2$ & $n+1$  \\
  $\surd$  & $-$ & det & $2n+3$ &  $2n+2$  \\
 $\surd$ & $\surd$ & det & $2n+3$ & \ \\
\hline \multicolumn{5}{l}{}  \\
\multicolumn{5}{l}{New Results (main cache of arbitrary size $k$):}  \\
\hline   Bypass & Reorg' & det'/rand'  & Upper Bound & Lower Bound   \\ \hline
  $-$ & $\surd$    & det & $(n+1)(k+1)-1$ & $(n+1)(k+1)-1$  \\
  $-/\surd$ & $-/\surd$    & det & $O(nk)$ & $\Omega(nk)$  \\
  $-/ \surd$ & $- /\surd$ & rand & $O(\log k \log n)$ & $\Omega(\log k + \log n)$
 \\\hline
\end{tabular}
\end{center}
\caption{Summary of the results in \cite{BETW01} and in this paper,
for the $(n,k)$ companion cache.}
\label{table:results}
\end{table}

\paragraph*{Our Results:}
This paper studies deterministic and randomized caching algorithms for a
$(n,k)$-companion cache. We consider the version where reorganization is
allowed but bypassing is not.
We show that the deterministic competitive ratio is exactly $(n+1)(k+1)-1$. 
For randomized algorithms, we present an upper bound of $O( \log n \log k)$ on
the competitive ratio, and a lower bound of $\Omega(\log n + \log k)$. For
the special case of $k=1$ that was studied in \cite{BETW01}, our bounds on
the randomized competitive ratio are tight up to a constant factor. The
results of \cite{BETW01} and those of this paper are summarized and compared
in Table~\ref{table:results}.

We note that any algorithm for the reorganization model can be implemented
(in online fashion) in the no-reorganization model while incurring a cost at
most two times larger, and any algorithm for the bypassing model can be
implemented (in online fashion) in the no-bypassing model while incurring a
cost at most two times larger. Thus, the competitive ratio (both randomized
and deterministic) differs by at most a constant factor between the different
models.

The techniques we use generalize  \emph{phase partitioning} and
\emph{marking algorithms}~\cite{KMRS88,FKLMSY91}.

\section{The Problem}

In the $(n,k)$-companion caching problem, there is a slow {\em
main memory} and a fast {\em cache}. The items in main memory are
partitioned into $m$ {\em types}, the set of types is $T$ ($|T|=m$).
The cache consists of a two separate components:
\begin{itemize}
\item The Main Cache: Consisting of a cache of size $k$ for each type.
\emph{I.e.}, every type $t$, $1\leq t \leq m$, has its own cache of size $k$
which can hold only items of type $t$.
\item The Companion Cache: A cache of size $n$ which can hold items of any type.
\end{itemize}
We refer to these components collectively simply as {\em the cache}. If an
item is stored somewhere in the cache, we say it is {\em cached}. Our basic
assumptions are that there are at least $k+1$ items of every type and that
the number of types, $m$, is greater than the size of the companion cache,
$n$.

A caching algorithm is faced with a sequence of requests for items. When an
item is requested it must be cached (\ie, bypassing is not allowed). If the
item is not cached, a fault occurs. The goal  is to minimize the number of
faults. A caching algorithm can swap items of the same type between the main
and companion caches without incurring any additional cost (\ie,
reorganization is allowed).

We use the competitive ratio to measure the performance of online
algorithms. Formally, given an item request sequence $\sigma$, the
cost of an online algorithm $A$ on $\sigma$, denoted by
$\cost_A(\sigma)$, is the number of faults incurred by $A$. An
algorithm is called $r$-competitive if there exists a constant $c$, such
that for any request sequence $\sigma$,
 \[ E[\cost_A(\sigma)] \leq r \cdot \cost_{\opt}(\sigma) +c .\]

To simplify the analysis later, we mention the following fact (attributed to
folklore):
\begin{proposition}
We may assume that {\opt} is lazy, {\sl i.e.}, {\opt} evicts an item only
when a requested item is not cached.
\end{proposition}

\section{Lower bounds on the competitive ratio}

Straightforward lower bounds follow from the classical paging problem.

\begin{theorem}
The deterministic competitive ratio for the $(n,k)$-companion caching
problem is at least $(n+1)(k+1)-1$. The randomized competitive ratio is at
least  $H_{(k+1)(n+1)-1}=\Omega(\log n + \log k)$.
\end{theorem}
\begin{proof}
Consider the situation where there are $(n+1)(k+1)$ items of $n+1$
types, $k+1$ items of each type. In this case, a caching algorithm
has $(n+1)k+n=(n+1)(k+1)-1$ cache slots available. Comparing this
situation to the regular paging problem with a main memory of
$(n+1)(k+1)$ items and a cache size of $(n+1)(k+1)-1$, we find the
two problems are exactly the same. A companion caching algorithm
induces a paging algorithm, and the opposite is also true. Hence a
lower bound on the competitive ratio for paging implies the same
lower bound for companion caching. We conclude there are lower
bounds of $(n+1)(k+1)-1$ on the deterministic competitive ratio
and $H_{(n+1)(k+1)-1}= \Omega(\log n+ \log k)$ on the randomized
competitive ratio for companion caching.
\end{proof}

\section{Phase Partitioning of Request Sequences} \label{sec:phase}

In \cite{KMRS88,FKLMSY91} the request sequence for the paging problem is
partitioned into \emph{phases} as follows: A phase begins either at the
beginning of the sequence or immediately after the end of the previous
phase. A phase ends either at the end of the sequence or immediately before
the request for the $(k+1)$st distinct page in the phase. Similarly, we
partition the request sequence for the companion caching problem into
{phases}. However, the more complex nature of our problem implies  more
complex partition rules.

\newcommand{\leti}{\leftarrow}
\begin{figure}[htb]
\begin{center}
\fbox{
\begin{minipage}{4in}
\begin{tabbing}
\ \ \ \ \ \ \= \ \ \ \ \ \ \=  \ \ \ \ \ \ \= \ \ \ \ \ \ \ \= \kill
 $P_i$: \> \> The indices of the requests \emph{associated with} phase $i$.\\
 $D_i$: \> \> The indices of the requests issued \emph{during} phase $i$.\\
 $N(t)$: \> \> The indices of  requests of type $t$ that have not yet \\
 \>\>been associated with a phase.\\
  $M(t)$ \> \> $= \{\sigma_\ell | \ell\in N(t)\}$\\

  \\
 For every type $t\in T$: $M(t)\leti \emptyset$, \ $N(t)\leti
 \emptyset$ \\
 $P_1\leti \emptyset$, \  $D_1\leti \emptyset$ \\
 $i\leti 1$\\
 For $\ell\leti 1,2,\ldots$ \ \ \ \emph{Loop on the requests} \+ \\
 Let $\sigma_\ell$ be the current request and $t_0$ be its type.\\
 Let $m_t \leti \begin{cases} \max\{0, |M(t)|-k\} & t \neq t_0 \\
                    \max\{0, |M(t_0)\cup \{\sigma_\ell\}|-k\} & t = t_0
                    \end{cases}$\\
 If $\sum_{t \in T}m_t>n$ then \ \ \
   \emph{End of Phase Processing:} \+ \\
   For every type $t\in T$ such that $m_t>0$ do \+ \\
   $P_i \leti P_i \cup N(t)$ \\
   $M(t) \leti \emptyset$, \
   $N(t) \leti \emptyset$ \- \\

   $i \leti i+1$  \\
    $P_i \leti \emptyset$, \
    $D_{i}\leti \emptyset$ \-  \\
 $D_i\leti D_{i} \cup \{\ell\}$\\
 $N(t_0)\leti N(t_0) \cup\{\ell\}$\\
 $M(t_0)\leti M(t_0) \cup\{\sigma_\ell\}$\-
\end{tabbing}
\end{minipage}}
\end{center}
 \caption{ Phase partition rules described as an algorithm.}
\label{fig:phase-partition}
\end{figure}

Let  $\sigma = \sigma_1, \sigma_2, \ldots, \sigma_{|\sigma|}$
denote the request sequence. The indices of the sequence are
partitioned into a sequence of disjoint consecutive subsequences
$D_1, D_2, \ldots, D_f$, whose concatenation gives
$\{1,\ldots,|\sigma|\}$. The indices are also partitioned into a
sequence of disjoint (ascending) subsequences $P_1, P_2, \ldots,
P_f$.

In Figure \ref{fig:phase-partition} we describe how to generate
the sequences $D_i$ and $P_i$. $D_i$ is a consecutive sequence of
indices of requests {\sl issued during} phase $i$. $P_i$ is a
(possibly non-consecutive, ascending) sequence of indices of
requests {\sl associated} with phase $i$. Note that $\ell \in D_i$
does not necessarily imply that $\ell\in P_i$ and vice versa. What
is true is that $\ell\in D_i$ implies either that $\ell \in
P_{i'}$ for some $i' \geq i$, or $\ell\notin P_{i'}$ for all $i'$.
Note also that for all $i$, $\max D_i \geq \max P_i$.

Given a set of indices $A$ we denote by $\items(A)=\{\sigma_\ell |
\ell\in A\}$ the set of items requested in $A$, and by $\types(A)$
the set of types of items in $\items(A)$.

Table~\ref{table:phase-example} shows an example of phase partitioning.

\begin{table}
\begin{center}
\begin{tabular}{|l| *{10}{p{1pt}} *{8}{p{1pt}} *{5}{p{3pt}} *{2}{p{2pt}}|}
\hline
           & \multicolumn{9}{r}{ }& 1& 1 &1 & 1& 1& 1& 1& 1&1 & 1 & 2 & 2 & 2& 2&2 & \\
Indices    & 1     & 2     & 3& 4& 5& 6 & 7 & 8 & 9 & 0 &1 & 2 & 3 & 4 & 5 &
             6 & 7 & 8 & 9 & 0 & 1 & 2 & 3 & 4 & \\
\hline
 Req. seq. & $a_1$ & $b_1$ & $d_1$ & $c_1$ & $a_2$ & $a_3$ & $b_2$ & $a_4$ & $b_3$ & $c_2$
           & $b_4$ & $a_5$ & $c_3$ & $d_2$ & $b_1$ & $c_4$ & $a_3$ & $a_2$
           & $a_1$ & $a_3$ & $b_2$ & $b_3$ & $b_5$ & $d_3$ &$ \ldots$\\
           \hline
Phase &     \multicolumn{10}{c}{$i=1$}  \vline &
\multicolumn{8}{c}{$i=2$}\vline &
            \multicolumn{5}{c}{$i=3$} \vline & \multicolumn{2}{c}{ }\vline  \\ \hline
           $D_i$ & \multicolumn{10}{c}{$\{1,\ldots,10\}$} \vline&
                    \multicolumn{8}{c}{$\{11,\ldots,18\}$} \vline&
                     \multicolumn{5}{c}{$\{19,\ldots,23\}$} \vline& & \ \\ \hline
          $P_i$ & \multicolumn{10}{c}{$\{ 1, 2, 5, 6, 7, 8, 9\}$} \vline &
          \multicolumn{8}{c}{$\biggl\{ \begin{matrix} 4,10, 12, 13, \\ 16,17,18\end{matrix}\biggr\}$} \vline &
          \multicolumn{5}{c}{$\biggl\{\begin{matrix}3,11,14,15, \\ 21,22,23 \end{matrix} \biggr\}$} \vline &  & \\ \hline
$\types(P_i)$ & \multicolumn{10}{c}{$\{a,b\}$} \vline &
                \multicolumn{8}{c}{$\{a,c\}$} \vline &
                \multicolumn{5}{c}{$\{ b,d \}$} \vline & & \\
\hline
\end{tabular}
\end{center}
\caption{An example for  an $(n,k)$-companion
caching problem where $n=3$ and $k=2$. The types are denoted by
the letters $a,b,c,d$. The $i$th item of type $\beta \in \{a,b,c,d\}$ is denoted by
$\beta_i$. Note that the requests for
items $d_1$ and $d_2$ in this example are in $P_3$,  even though
of they are issued \emph{during} phases~1 and 2 (i.e., belong
to $D_1$ and $D_2$).} \label{table:phase-example}
\end{table}

In \cite{KMRS88} it is shown that any paging algorithm faults at least once
in each complete phase.
Here we show a similar claim for companion caching.

\begin{proposition} \label{cl:opt_lb}
For any (online or offline) caching algorithm, it is possible to associate with each
phase (except maybe the last one) a distinct fault.
\end{proposition}
\begin{proof}
Consider the request indices in $P_i$  together with the index $j$ that ends
the phase (\ie, $j=\min D_{i+1}$). One of the items in $\items(P_i)$ must be
evicted after being requested and before $\sigma_j$ is served. This is
simply because the cache cannot hold all these items simultaneously. We
associate this eviction with the phase.

We must show that we have not associated the same eviction to two distinct
phases. Let ${i_1}$ and ${i_2}$ be two distinct phases, $i_1<i_2$. If the
evictions associated with ${i_1}$ and ${i_2}$ are of different items then they
are obviously distinct. Otherwise, the evictions associated with ${i_1}$ and
${i_2}$ are of the same type $t$, and $t\in\types(P_{i_1})\cap
\types(P_{i_2})$, which means that all indices $\ell \in P_{i_2}$, where $\sigma_\ell $ is
of type $t$, must have $\ell > \max D_{i_1} $. Thus, an eviction associated with
phase ${i_2}$ cannot be associated with phase ${i_1}$.
\end{proof}

To help clarify our argument in the proof of Proposition~\ref{cl:opt_lb},
consider the third phase in Table~\ref{table:phase-example}. Here
$\items(P_3)=\{b_1,b_2,b_2,b_4,b_5,d_1,d_2\}$, and the phase ends because of the
request to $d_3$. It is not possible that all these items reside in the cache
simultaneously and thus at least one of the items in $\items(P_3)$ must be
evicted before or on the request for item $d_3$. The item evicted can be either
some $b_i$, $i\in\{1,2,3,4,5\}$, or some $d_i$, $i\in\{1,2\}$. If, for example, the item
evicted is some $b_i$, then this eviction must have occurred after $\max D_1$
--- the end of the first phase --- and therefore it cannot be an eviction
associated with the first phase.

\section{{Deterministic Marking} Algorithms}

In a manner similar to \cite{KMRS88}, based on the phase partitioning of
Section~\ref{sec:phase}, we define a class of online algorithms called
\emph{marking} algorithms.

\begin{definition}
During the request sequence an item $e\in \bigcup_t M(t)$ is
called \emph{marked} (see Figure~\ref{fig:phase-partition} for a
definition of $M(t)$). An online caching algorithm that never
evicts marked items is called a \emph{marking algorithm}.
\end{definition}

\noindent Remarks: \begin{enumerate} \item The phase partitioning and dynamic
update of the set of marked items can be performed in an online fashion (as
given in the algorithm of Fig.~\ref{fig:phase-partition}). \item At any point
in time, the cache can accommodate all marked items. \item Unlike the marking
algorithms of \cite{KMRS88}, it is not true that immediately after $\max D_i$
all marks of the $i$th phase are erased. Only the marked items of types $t\in
\types(P_i)$ will have their markings erased immediately after $\max D_i$.
\end{enumerate}

For a specific algorithm, at any point in time during the request sequence,
a type $t$ that has more than $k$ items in the cache is called
\emph{represented in the companion cache}. Note that for marking algorithms,
a type is in $\types(P_i)$ if and only if it is represented in the companion
cache at  $\max D_i$ or it is the type of the item that ended phase $i$.

\begin{proposition} \label{cl:mark_ub}
The number of faults of any marking algorithm on requests whose indices are in
$P_i$ is at most $n(k+1)+k= (n+1)(k+1)-1$.
\end{proposition}
\begin{proof}
Each item $e$ of type $t$ requested in request index $\ell \in P_i$, is marked
and is not evicted until after $\max D_{i}$. We note that $|\types(P_{i})|\leq
n+1$ since at most $n$ types are represented in the companion cache, and the
type of the item whose request ends the phase may also be in $\types(P_i)$.
Thus, $|\items(P_i)|\leq (n+1)k+n$.
\end{proof}

We conclude from Proposition~\ref{cl:mark_ub} and
Proposition~\ref{cl:opt_lb}:

\begin{theorem}
Any {marking} algorithm is  $(n+1)(k+1)-1$ competitive.
\end{theorem}
\begin{proof}
Immediate from Proposition~\ref{cl:mark_ub} and Proposition~\ref{cl:opt_lb}.
\end{proof}

Since the marking property can be realized by deterministic algorithms, we
conclude

\begin{corollary}
The deterministic competitive ratio of the\/ $(n,k)$-companion caching problem
is $(n+1)(k+1)-1$.
\end{corollary}

\section{Randomized {Marking} Algorithms}

In this section we present an $O(\log n \log k)$ competitive randomized marking
algorithm.
The building blocks of our randomized algorithms are the following three eviction strategies:

On a fault on an item of type $t$:
\begin{description}
\item [Type Eviction.] Evict an item chosen uniformly at random
among all unmarked items of type $t$ in the cache.

\item [Cache-wide Eviction.] Let $T$ be the set of types represented in
the companion cache, let $U$ be the set of all unmarked items
in the cache
whose type is in $T \cup \{t\}$.
Evict an item chosen uniformly at random from $U$.

\item [Skewed cache-wide eviction.] Let $T$ be the set of types represented in the companion cache, let $T' \subset T \cup \{t\}$ be the set of types
with at least one unmarked item in the cache. Choose $t'$ uniformly at random from $T'$, let $U$ be the set of all unmarked items of type $t'$,
and evict an item chosen uniformly at random from $U$.
\end{description}

Remarks:

\begin{enumerate}
\item  Type eviction may not be
possible as there may be no unmarked items of type $t$ in the
cache.
\item
Cache-wide eviction and skewed cache-wide eviction are always possible, if
there are no unmarked pages of types represented in the companion cache and no
unmarked pages of type $t$ in the cache then the fault would have ended the
phase.
\end{enumerate}

\noindent  The algorithms we use are:

\paragraph*{Algorithm {\marke}.}
Given a request for item $e$ of type $t$, not in the cache: Update all phase
related status variables (as in the algorithm of Figure
\ref{fig:phase-partition}).
\begin{itemize}
\item If $t$ is not represented in the companion cache and there are unmarked
items of type $t$, use type-eviction.

\item Otherwise --- use cache-wide eviction.
\end{itemize}

\paragraph*{Algorithm {\markd}.}
Given a request for item $e$ of type $t$, not in the cache: Update all phase
related status variables (as in the algorithm of Figure
\ref{fig:phase-partition}). Let the current request index be $j\in D_i$, $i\geq
1$.
\begin{itemize}
\item If $t$ is not represented in the companion cache and there are unmarked
items of type $t$, use type-eviction.

\item If $t$ is represented in the companion cache, $e \in \items(P_{i-1})$, and there are unmarked items of type $t$,
use type eviction.

\item Otherwise --- use skewed cache-wide eviction.
\end{itemize}

\paragraph*{Algorithm \textsf{TP}.} If $k<n$ use {\marke}, otherwise, use
{\markd}.

\medskip

\noindent In the rest of this section we prove: 

\begin{theorem} \label{thm:lognlogk}
Algorithm \textsf{TP} is $O(\log n \log k)$ competitive.
\end{theorem}

\subsection{Basic Definitions and Proof Overview}
\label{subsec:basic}

We give an analogue to the definitions of new and stale pages used in the
analysis of the randomized marking paging algorithm of \cite{FKLMSY91}.

\begin{definition}
For phase $i$ and type $t$, denote by $i^{-t}$ the largest index $j<i$ such
that $t \in \types(P_j)$. If no such $j$ exists we denote $i^{-t}=0$, and use
the convention that $P_0=\emptyset$. Similarly, $i^{+t}$ is the smallest index
$j> i$ such that $t \in \types(P_j)$. If no such index exists, we set
$i^{+t}=``\infty"$, and use the convention that $P_\infty=\emptyset$.
\end{definition}

\begin{definition}
\label{def:new} An item $e$ of type $t$ is called \emph{new} in $P_i$ if $e\in
\items(P_i)\setminus\items(P_{i^{-t}})$.  We denote by $g_{t,i}$ the number of
new items of type $t$ in $P_i$. Note that if $t\notin \types(P_i)$ then
$g_{t,i}=0$.
\end{definition}

Let $i_{\text{end}}$ denote the index of the last \emph{completed}
phase.

\begin{definition}
For $t\in \types( P_i)$, let $L_{t,i}=\items(P_i) \cap \{$items of
type $t\}$. Note that $|L_{t,i}| \geq k$. Define
\[ \ell_{t,i}= \begin{cases} |L_{t,i}| -k & i< i_{\text{end}}
\; \land \; t\in \types(P_i) \setminus \types(P_{i+1}), \\
0 & \text{otherwise.}
\end{cases} \]
\end{definition}

We will use the above definitions to give an amortized lower bound
(see Lemma \ref{lem:opt_lb})  on the cost of {\opt} of dealing
with the sequence $\sigma$:
\begin{equation} \cost_{\opt}(\sigma) \geq \tfrac{1}{4} \sum _{i\leq i_{\mathrm{end}}}
\sum _{t \in P_i} (g_{t,i}+ \ell_{t,i^{-t}}). \label{eq:optlb} \end{equation}

Our algorithms belong to a restricted
family of randomized algorithms, specifically \emph{uniform type
preference} algorithms defined below. The main advantage of using
such algorithms is that their analysis is simplified as they have
the property that while dealing with requests $\sigma_j$, $j\in
D_i$, the companion cache is restricted to containing only items
of types in $\types(P_{i}) \cup \types(P_{i-1})$.

\begin{definition}
A \emph{type preference} algorithm is a marking algorithm such that when a fault
occurs on an item of a type that is not represented in the companion cache, it
evicts an item of the same type, if this is possible.
\end{definition}

\begin{definition}
A \emph{uniform type preference} algorithm is a randomized type preference
algorithm maintaining the invariant that at any point in time between request
indices $1+\max D_{i^{-t}}$ and $\max D_i$, inclusive, and any type $t\in
\types(P_i)$, all unmarked items of type $t$ in $\items(P_{i^{-t}})$ are
equally likely to be in the cache.
\end{definition}

Note that both {\marke} and {\markd} are uniform type preference algorithms.

We use a charge-based amortized analysis to compute the online cost of
dealing with a request sequence $\sigma$. We charge the expected cost of all
but a constant number of requests in $\sigma$ to at least one of two
``charge counts", $\mbox{charge}(D_i)$ and/or $\mbox{charge}(P_j)$ for some
$1 \leq i\leq j \leq i_{\text{end}}$. The total cost associated with the
online algorithm is bounded above by a constant times $\sum_{1\leq i \leq i_{\text{end}}}
\mbox{charge}(D_i) + \sum_{1\leq i \leq i_{\text{end}}} \mbox{charge}(P_i)$,
excluding a constant number of requests.

Other than a constant number of requests, every request
$\sigma_\ell\in \sigma$ has $\ell \in D_{i_1}\cup P_{i_2}$ for
some $1 \leq i_1 \leq i_2  \leq i_{\text{end}}$.

We use the following strategy to charge the cost associated with this request
to one (or more) of the $\mbox{charge}(D_i)$, $\mbox{charge}(P_j)$:
\begin{enumerate}

\item If $\ell \in P_i$ and $\mbox{type}(\sigma_\ell) \in
\types(P_i)\setminus \types(P_{i-1})$ then we charge the (expected) cost of
$\sigma_\ell$ to $\mbox{charge}(P_i)$. These charges can be amortized against
the cost of {\opt} to deal with $\sigma_\ell$. This amortization is summarized
in Proposition \ref{prop:markd_new_ub} (for any uniform type preference
algorithm).

\item If $\ell \in D_i$ and $\mbox{type}(\sigma_\ell) \in \types(P_{i-1})$ then
we charge the (expected) cost of $\sigma_\ell$ to
$\mbox{charge}(D_i)$. These charges will be amortized against the
cost of {\opt} to within a poly-logarithmic factor. This
amortization is summarized in Proposition \ref{prop:marke_old_ub}
for algorithm {\marke} and Proposition \ref{prop:markd_old_ub} for
algorithm {\markd}.
\end{enumerate}

To compute the expected cost of a request $\sigma_\ell$, $\ell \in D_i$,
$\mbox{type}(\sigma_\ell) \in \types(P_{i-1})$, we introduce an analogue to the
concept of ``holes" used in~\cite{FKLMSY91}. In \cite{FKLMSY91} holes were defined to be
stale pages that were evicted from the cache.

\begin{definition}\label{def:hi}
We define the number of holes during $D_i$, $h_i$, to be the maximum over
the indices  $j\in D_i$ of the total number of items of types in
$\types(P_{i-1})$ that were requested in $P_{i-1}$ but are not cached when
the $j$th request is issued.
\end{definition}

\subsection{Analysis of the Competitive Ratio for Algorithm \textsf{TP}}
\label{subsec:detail}

\subsubsection{Lower Bounds on \opt}

\begin{proposition} \label{prop:lb1}
For any request sequence $\sigma$,
\[ \cost_{\opt}(\sigma) \geq   \sfrac 1 2\sum _{i\leq i_{\mathrm{end}}} \; \sum _{t } g_{t,i}  \]
\end{proposition}
\begin{proof}
We may assume without loss of generality that {\opt} is lazy.
Let $C_i$ be the items in {\opt}'s cache at the end of $P_i$
($C_0=\emptyset$). For pairs $i,t$, let $G'_{t,i}$ be the set of new items
in $\items(P_i)$ of type $t$  that do not appear in $C_{i^{-t}}$, and let
$G''_{t,i}$ be the set of new items in $\items(P_i)$ of type $t$ that
\emph{do} appear in $C_{i^{-t}}$. {From} the definitions,
$|G'_{t,i}|+|G''_{t,i}|=g_{t,i}$.

First we show that $\cost_{\opt}(\sigma) \geq \sum_i \sum_{t\in \types(P_i)}
|G'_{t,i}|$. For any $t\in \types(P_i)$ and for any item $a\in G'_{t,i}$ we
have $a\in
\items(P_i)\setminus C_{i^{-t}}$. Thus, for any lazy algorithm, the first
request for $a$ in $P_i$ is a fault. Let the request sequence
$\sigma=\sigma_1, \sigma_2, \ldots$, we define
\begin{align*}
J(G'_{i,t}) &= \{j | j=\min \{\ell|\sigma_\ell =a, \ell\in P_i\}, a\in
G'_{i,t}\}, \text{ and } & J'_i &= \cup_{t \in \types(P_i)} J(G'_{i,t}).
\end{align*}

For any lazy algorithm, $J'_i$ is a set of request indices that result in
faults. We are left to argue that $J'_{i_1}\cap J'_{i_2} = \emptyset$ for
$i_1\neq i_2$, but this is obvious, since $J'_i\subseteq P_i$, and
$P_{i_1}\cap P_{i_2}=\emptyset$.

Next, we show that $\cost_{\opt} (\sigma) \geq \sum_i \sum_{t\in \types(P_i)}
|G''_{t,i}|$. Note that $G''_{t,i^{+t}}\subseteq C_i$, \ie, items in
$G''_{t,i^{+t}}$ are in {\opt}'s cache after serving $\max D_i$. As {\opt} is
lazy, all items in $G''_{t,i^{+t}}$ must reside in the cache continuously
since request index $\max{D_{i^{-t}}}$. The slots used to store these items
will be unavailable to deal with requests whose indices are in $P_i$.
Consider the behavior of {\opt} on the request sequence $\sigma$. We claim
that {\opt} must have at least $\sum_{t\in \types(P_i)} |G''_{t,i^{+t}}|$
evictions of items that were requested in $P_i$, \emph{after} their request,
and before $\max D_i$.

For every type $t \in \types(P_i)$ there were $k+\alpha_t$ requests to
different items of type $t$ in $P_i$, $\sum_{t\in \types(P_i)} \alpha_t =
n$. The total memory that we have available to deal with these
$n+k|\types(P_i)|$ different items is no more than $n+k|\types(P_i)|$ minus
the number of slots that are unavailable, {\sl i.e.}, the number of slots
available for requests whose indices are in $P_i$ is no more than
$$n+k|\types(P_i)| - \sum_{t\in \types(P_i)} |G''_{t,i^{+t}}|.$$ Thus, {\opt}
must have evicted at least $\sum_{t\in \types(P_i)} |G''_{t,i^{+t}}|$ of them
by the end of $\max D_i$.

To argue that we do not count the items in $G''_{t,i^{+t}}$ more than once,
we note that if $t\in \types(P_{i_1}) \cap \types(P_{i_2})$ for $i_1 \neq
i_2$ then $i_1^{+t} \neq i_2^{+t}$.
\end{proof}

\begin{proposition} \label{prop:lb2}
For any request sequence $\sigma$,
\[ \cost_{\opt}(\sigma) \geq
\sfrac 1 2 \sum _{i< i_{\mathrm{end}}} \; \sum _{t } \ell_{t,i} .
\]
\end{proposition}
\begin{proof}
Once again, we can assume {\opt} is lazy. Fix $i<i_{\text{end}}$, and a type
$t\in \types(P_i)$ such that $t \notin P_{i+1}$. Let
\begin{align*}
L'_{t,i} &= L_{t,i} \setminus C_{i+1}; & L''_{t,i} &= L_{t,i} \cap C_{i+1}.
\end{align*}
Every item $e \in L'_{t,i}$ has some $\ell \in P_i$ such that $\sigma_\ell =
e$, and $e$ was evicted by {\opt} later (but before $\max D_{i+1}$). Let
$\tilde{\ell}$ be the largest such $\ell\in P_i$. We associate one eviction
of $e$ with index $\tilde{\ell}$. 
In this way every eviction is associated with at most
one index.
Indeed, the associated eviction occurs not before $\min D_i$,
and before $\max D_{i+1}$. At this time frame, only items from
$L_{t,i}$ and $L_{t,i+1}$ could have been associated with this eviction,
but $t \notin \types(P_{i+1})$.
Therefore $\cost_{\opt}(\sigma) \geq \sum_i
\sum_{t \in \types(P_i)} |L'_{t,i}|$.

Let $t\in \types(P_i)\setminus \types(P_{i+1})$, and assume $|L''_{t,i}| >
k$. Such items occupy at least $$\phi_{i+1}=\sum_{t \in
\types(P_i)\setminus\types(P_{i+1})}\max\{|L''_{t,i}|-k,0\}$$ slots in the
companion cache at time $\max D_{i+1}$. In $P_{i+1}$ there are requests for
$|\types(P_{i+1})|k+ n$ different items, but considering the cache at time
$\max D_{i+1}$, these items occupy at most $|\types(P_{i+1})|k+ n-
\phi_{i+1}$ slots. This means that at least $\phi_{i+1}$ of the items
$\items(P_{i+1})$, were evicted subsequently to being requested at request
indices in $P_{i+1}$ and no later than $\max D_{i+1}$.

Associate each such eviction of item $a \in
\items(P_{i+1})$ with the largest  index $\ell\in P_{i+1}$ such that
$\sigma_\ell=a$. Note that each such eviction is associated with only one
index, and therefore
\[ \cost_{\opt}(\sigma) \geq \sum_{i<i_{\text{end}}} \;
\sum_{t \in \types(P_i) \setminus \types(P_{i+1})}\max\{|L''_{t,i}|-k,0\}.
   \]
We conclude
\begin{eqnarray*}
\cost_{\opt}(\sigma) &\geq&
 \max \Bigl \{ \sum_{i<i_{\text{end}}} \sum_{t \in \types(P_i)} |L'_{t,i}| ,
  \sum_{i< i_{\text{end}}}\sum_{\substack{t \in \types(P_i), \\ t \notin
\types(P_{i+1})}}\max\{|L''_{t,i}|-k,0\} \Bigr \} \\
 &\geq& \tfrac{1}{2} \sum_{i<i_{\text{end}}}
\sum_{\substack{t \in \types(P_i), \\ t \notin \types(P_{i+1})}} \max\{
|L'_{t,i}| + |L''_{t,i}|-k,0\} = \tfrac{1}{3}\sum_{i<i_{\text{end}}}
\ell_{t,i}. \qquad \qed
\end{eqnarray*}
\renewcommand{\qed}{}
\end{proof}

By taking a convex combination of the lower bounds of
Proposition~\ref{prop:lb1} and Proposition~\ref{prop:lb2}, and by algebraic
manipulations, we conclude:
\begin{lemma} \label{lem:opt_lb}
For any request sequence $\sigma$,
\[ \cost_{\opt}(\sigma)
\geq \tfrac{1}{4} \sum _{i\leq i_{\mathrm{end}}} \sum _{t \in P_i}
(g_{t,i}+ \ell_{t,i^{-t}}). \]
\end{lemma}


\subsubsection{Upper Bounds on \textsf{TP}}


\begin{proposition} \label{prop:holes}
Consider a marking algorithm, a phase $i$, a type $t\in
\types(P_i)\setminus\types(P_{i-1})$, and a request index $\max
D_{i^{-t}}<j\leq \max D_i$. Let $H$ be the set of items of type $t$ that were
requested in $P_{i^{-t}}$ and evicted afterward without being requested
again, up to request index $j$ (inclusive). Then $|H| \leq \hat{g}_{t,i}+
\ell_{t,i^{-t}}$, where $\hat{g}_{t,i}\leq g_{t,i}$ is the number of new
items of type $t$ requested after $\max D_{i^{-t}}$ and up to time $j$
(inclusive).
\end{proposition}
\begin{proof}
Recall that $L_{t,i^{-t}}$ is the set of marked items of type $t$ after
serving $\max D_{i^{-t}}$. Let $\hat{G}_{t,i}$ be the set of items requested
after request $\max D_{i^{-t}}$ and before request index  $j$ that are not in
$L_{t,i^{-t}}$, \ie, $\hat{G}_{t,i}$ is the set new items of type $t$
requested up to request index $j$.

If $i^{-t}=0$ then $H\subseteq L_{t,i^{-t}}=\emptyset$. Otherwise, as
$H\subseteq L_{t,i^{-t}}\subseteq L_{t,i^{-t}} \cup \hat{G}_{t,i}$, and $k$
items of $L_{t,i^{-t}} \cup \hat{G}_{t,i}$ are always in the (main) cache,
we conclude

\[ |H| \leq | L_{t,i^{-t}} \cup \hat{G}_{t,i}| -k \leq (|L_{t,i^{-t}}| -k ) + |\hat{G}_{t,i}| =
 \ell_{t,i^{-t}} + \hat{g}_{t,i} . \qquad \qed\]
 \renewcommand{\qed}{}
\end{proof}

\begin{proposition} \label{prop:markd_new_ub}
For a uniform type preference  algorithm, the expected number of faults on
request indices in $P_i$ for items of type  $t \in \types(P_i) \setminus
\types(P_{i-1})$ is at most \( (1+H_{n+k})(g_{t,i}+ \ell_{t,i^{-t}}) . \)
\emph{I.e.}, $\text{charge}(P_i)\leq (1+H_{n+k})(g_{t,i}+ \ell_{t,i^{-t}})$.
\end{proposition}
\begin{proof}
Fix a  type $t\in \types(P_i) \setminus \types(P_{i-1})$. There are $g_{t,i}$
faults on new items of type $t$, the rest of the faults are on items in
$L_{t,i^{-t}}$ that were evicted before being requested again. By
Proposition~\ref{prop:holes}, the number of items in $L_{t,i^{-t}}$ that are
not in the cache at any point of time is at most
$\hat{g}_{t,i}+\ell_{t,i^{-t}}\leq g_{t,i}+\ell_{t,i^{-t}}$. For any $a$, $b$
in $L_{t,i^{-t}}$ that have not been requested after $\max D_{i^{-t}}$, the
probability that $a$ has been evicted since $1+ \max D_{i^{-t}}$ is equal to
the probability that $b$ has been evicted since $1+ \max D_{i^{-t}}$.

Let $r$ denote the number of items in $L_{t,i^{-t}}$ that have been requested
after $\max D_{i^{-t}}$. There are $|L_{t,i^{-t}}| -r $ unmarked items
of $L_{t,i^{-t}}$. The probability that an unmarked item of $L_{t,i^{-t}}$ is 
not cached
is therefore at most $(g_{t,i}+\ell_{t,i^{-t}})/(|L _{t,i^{-t}}|-r )$. Thus, the
expected number of faults on requests indices in $P_i$ for items in
$L_{t,i^{-t}}$ is at most
\[
\sum_{r=0}^{|L_{t,i^{-t}}|-1} \frac{g_{t,i}+\ell_{t,i^{-t}}}{|L_{t,i^{-t}}|
 -r} \leq (g_{t,i}+\ell_{t,i^{-t}} ) H_{|L_{t,i^{-t}}|} \leq
 (g_{t,i}+\ell_{t,i^{-t}} ) H_{n+k}. \qquad \qed
\]
\renewcommand{\qed}{}
\end{proof}

The following proposition is immediate from the definitions.
\begin{proposition} \label{prop:nprop}
A type preference algorithm has the following properties:
\begin{enumerate}
\item During $D_i$, only types in $\types(P_{i-1}) \cup \types(P_i)$ may be
represented in the companion cache.
\item During $D_i$, when a type $t \in \types(P_{i}) \setminus \types(P_{i-1})$ becomes
 represented in the companion cache,
there are no unmarked cached  items of type $t$, and $t$ stays represented in
the companion cache until $\max D_i$, inclusive.
\end{enumerate}
\end{proposition}

Recall the definition of $h_i$, the ``number of holes during $D_i$" (Definition
\ref{def:hi}).
\begin{proposition} \label{prop:toth}
For a type preference algorithm,
\begin{equation} \label{eq:toth}
h_i\leq \sum_{\substack{t\in \types(P_i)\setminus
\types(P_{i-1})}} (g_{t,i}+ \ell_{t,i^{-t}})+
  \sum_{t \in \types(P_{i-1})} g_{t,(i-1)^{+t}} .
  \end{equation}
\end{proposition}
\begin{proof}
At time $\min D_i$, among the types in $\types(P_{i-1})\cup \types(P_{i})$,
only types in $\types(P_i)\setminus \types(P_{i-1})$ may have uncached items
from $L_{t,i^{-t}}$. By Proposition~\ref{prop:holes}, the number of such
items at the beginning of $P_i$ is at most $\sum_{t\in \types(P_i) \setminus
\types(P_{i-1})} (\hat{g}_{t,i}+ \ell_{t,i^{-t}})$, where $\hat{g}_{t,i}$ is
the number of new items of type $t$ requested until $\max D_{i-1}$
(inclusive).

Consider an eviction of an item of type in $\types(P_i)\cup \types(P_{i-1})$
during $D_i$. The eviction \emph{must be caused by a request to an item of
either the same type or a type represented in the companion cache}.   By
Proposition~\ref{prop:nprop}, the types represented in the companion cache
are in $\types(P_i)\cup \types(P_{i-1})$, and therefore the type of the
requested item is also in $\types(P_i)\cup \types(P_{i-1})$. If the
requested item is an item of $L_{t,i^{-t}}$, then the number of uncached
items from $L_{t,i^{-t}}$ has not changed. Otherwise, it is a new item and
thus the number of new items increases. In total, we have bounded $h_i$ as in
Inequality~\eqref{eq:toth}.
\end{proof}

\begin{definition}
At any point during $D_i$, call a type $t \in \types(P_{i-1})$ that
has unmarked items in the cache and is represented in the
companion cache an \emph{active type}. 
Call an unmarked item $e \in \items(P_{i-1})$ of an active type an \emph{active item}.
\end{definition}
Note that an active item may not be cached.

The following proposition is immediate from the definitions.
\begin{proposition} \label{prop:monoton}
The following properties hold for type preference algorithms:
\begin{enumerate}
\item During $D_i$, the set of active types is monotone decreasing w.r.t.  containment.
\item During $D_i$, the set of active items is monotone decreasing w.r.t. containment.
\end{enumerate}
\end{proposition}

\begin{proposition} \label{prop:marke_old_ub}
For {\marke}, $\text{charge}(D_i)$ --- The expected number of faults  on
request indices in $D_i$ to types in $\types(P_{i-1})$ --- is at most \( h_i
(1+ H_{k+1}(1+ H_{(n+1)(k+1)})). \)
\end{proposition}
\begin{proof}
First, we count the expected number of faults on items in
$\cup _{t\in\types(P_{i-1})} L_{t,i-1}$.
By Proposition~\ref{prop:monoton}, the set of active items
is monotone decreasing, where an item becomes inactive either by
being marked, or because its type is no longer represented in the
companion cache.
Let $\langle m_j\rangle_{j=1,\ldots,w}$, be the sequence of numbers of active
items indexed on the \emph{events}. An event is either when an active item is
requested, or when an active type $t$ becomes inactive by being no longer
represented in the companion cache (it is possible that one request generates
two events, one from each case).

If the $j$th  event is a request for active item, then
$m_{j+1}=m_{j}-1$. Otherwise, if the $j$th event is the event of type $t$
becoming inactive, and before that event there were  $b$ active
items of type $t$, then $m_{j+1}=m_{j}-b$.

In the first case, the expected cost of the request, conditioned
on $m_j$, is at most $h_i /m_j$.

In the second case, there are $b$ items of type $t$ that became inactive, each
had probability at most $\frac{h_i}{m_j}$ of not being in the cache at that moment.
This means that the expected number of items among the up-until now active
items of type $t$, that are not in the cache, at this point in time, is
at most $\frac{h_i b }{m_j}$.

Let $g_t$ denote the number of new items of $P_{(i-1)^{+t}}$ (Definition
\ref{def:new}) requested during $D_i$ ($g_t \leq g_{t,(i-1)^{+t}}$). After type
$t$ becomes inactive, the number of items among $L_{t,i-1}$ that are not in the
cache can increase only when a new item of type $t$ is requested. Therefore the
expected number of items among $L_{t,i-1}$ that are not in the cache, after the
$j$th event (the event when $t$ became inactive), is at most $\frac{h_i
b}{m_j}+g_t$.

 Because of the uniform type eviction property of {\marke}, the
probability that an item in $L_{t,i-1}$ is not in the cache is the expected
number of items among $L_{t,i-1}$, and not in the cache, divided by the number
of unmarked items among $L_{t,i-1}$, and therefore the expected number of
faults on items of $L_{t,i-1}$ after the $j$th event is at most
 \[\sum_{a=1}^b(\frac{h_i b}{m_j}+g_t) \cdot \frac{1}{a}=
(\frac{h_i b}{m_j}+g_t) H_b. \] Note that $b\leq k+1$, and $\sum_{t\in P_{i-1}
} g_t \leq h_i$, and so the expected number of faults on items $e\in \cup_{t\in
\types(P_{i-1})} L_{t,i-1}$, conditioned on the sequence $\langle m_j\rangle_j$
is at most
\begin{equation} \label{eq:marke_faults}
  h_i H_{k+1} + h_i \sum_{j} \frac {(m_j - m_{j-1}) H_{k+1}}{m_j}
\end{equation}

The sequence $\langle m_j\rangle_j$ is itself a random variable, but  we can
give an upper bound on the expected number of faults on items $e\in \cup_{t\in
\types(P_{i-1})} L_{t,i-1}$ by bounding the \emph{maximum} of
Eq.~\eqref{eq:marke_faults} over all feasible sequences $\langle m_j\rangle_j$.
The worst case for \eqref{eq:marke_faults} will be when $\langle m_j\rangle_j =
\langle (n+1)(k+1) -j\rangle_{j=1}^{(n+1)(k+1)-1}$. Thus,
\begin{equation*}
h_i H_{k+1} (1+  \sum_{j} \frac {(m_j - m_{j+1})}{m_j}) \leq
  h_i H_{k+1} (1+  H_{(n+1)(k+1)})
\end{equation*}

We are left to add faults on new items of types in $\types(P_{i-1})$.
There are at
most $\sum_{t\in \types(P_{i-1})} g_{t,(i-1)^{+t}} \leq h_i$ such faults.
\end{proof}

We conclude
\begin{lemma} \label{lem:marke}
{\marke} is $O(\log k \max \{\log n, \log k\})$
competitive.
\end{lemma}
\begin{proof}
Each fault is counted by either $\text{charge}(P_i)$ (Proposition
\ref{prop:markd_new_ub}) or $\text{charge}(D_i)$ (Proposition
\ref{prop:marke_old_ub}) (faults on request indices in $D_i$ for items of
type in $\types(P_{i-1})\setminus \types(P_i)$ are counted twice), and by
Lemma ~\ref{lem:opt_lb}, we have that the expected number of faults of
{\marke} is at most
\[ \bigl (5 ( 1+H_{n+k}) + 10( 1+ H_{k+1} (1+H_{(n+1)(k+1)})) \bigr )\cost_{\opt} .
\qquad \qed \]
\renewcommand{\qed}{}
\end{proof}

For algorithm {\markd}, we have similar arguments.

\begin{proposition} \label{prop:markd_old_ub}
For {\markd}, $\text{charge}(D_i)$ --- The expected number of faults on
request indices in $D_i$ to types in $\types(P_{i-1})$ --- is at most \( h_i
(1+ H_{n+k}(1+H_{n+1}) ). \)
\end{proposition}
\begin{proof}
Denote by $A$ the set of active types at some point in time during $D_i$. We
claim that conditioned on the set of active types $A$, for any two active
types $t_1,t_2\in A$, the expected number of items in $L_{t_1,i-1}$ that are
not currently in the cache is equal to the expected number of items of
$L_{t_2,i-1}$ that are currently not in the cache.

We prove this by induction on the length of the request sequence. Before
request index ${\min D_i}$, all items among the active types are in the
cache, and the claim trivially holds. A fault on an item of $L_{t,i-1}$, not
currently in the cache, of active type $t$, is served by type eviction and
therefore the number of items from $L_{t,i-1}$ and not in the cache does not
change. A fault on an item of a type not represented in the companion cache that
has unmarked items, is served by type eviction, and since that type is not
active, it does not change the numbers of active items not in the cache.

If type eviction is not used then the fault is served by increasing the
number of items not in the cache among the active types. In this case a
skewed cache-wide eviction is used, which chooses a page to evict in a two
stage process, first choosing an active type uniformly at random, and then
choosing to evict an unmarked page of that type uniformly at random.

Given an active type $t\in A$ at some point in time $\ell\in D_i$, we use the
following notation: $u_t$ denotes the number of items in $L_{t,i-1}$ and not
currently in the cache, and $r_t$ the number of items among $L_{t,i-1}$
requested so far. Note that $u_t$ is a random variable.
The probability that an active item of active type $t$ is not in the cache,
conditioned on $u_t=y$, is $\frac{y}{|L_{t,i-1}| - r_t}$. Thus, the
probability that an active item of type $t$ is not in the cache is $\sum_y
\frac{y}{|L_{t,i-1}|-r_t} \Pr[u_t=y]$. Note that $\sum_y y
\Pr[u_t=y]=E[u_t]$.

Recall that the expectations $E[u_t]$ are all equal for active types $t$.
Assuming there are $a$ active types and that type $t$ is active,
\begin{equation}E[u_t] = \frac{\# \mbox{ active items not in cache }}{a} \leq
\frac{h_i}{a}.\label{eq:eut_bound1}\end{equation} Let $b_t$ be the number of
active types immediately before type $t$ became inactive. Thus, $E[u_t]
\leq\frac{h_i}{b_t}$ immediately before type $t$ becomes inactive. As of this
point of time, $u_t$ could increase only if a new item of type $t$ is
requested.
We can therefore bound the value
$ E[u_t] \leq \frac{h_i}{b_t} + g_{t,(i-1)^{+t}},$
throughout $D_i$. The expected number of faults on items of type $t$ is at
most
\[ \sum_{r=0}^{|L_{t_a,i-1}|-1} \frac{\frac{h_i}{b_t}+g_{t,(i-1)^{+t}}} {|L_{t_a,i-1}|- r}=
\left(\frac{h_i}{b_t}+g_{t,(i-1)^{+t}}\right) H_{|L_{t,i-1}|} .\] Using the
facts that $|L_{t,i-1}|\leq n+k$, and $\sum_{t\in \types(P_{i-1})}
g_{t,(i-1)^{+t}}\leq h_i$, and summing over all $t\in \types(P_{i-1})$, the
expected number of faults on types in $\types(P_{i-1})$ is at most
\[ \sum _{t \in \types(P_{i-1})} \left(\frac{h_i}{b_t}+g_{t,(i-1)^{+t}}\right) H_{|L_{t,i-1}|}
 \leq h_i  H_{n+k} \sum_{t \in \types(P_{i-1})} b_t^{-1} + h_i
 H_{n+k} \leq h_i  H_{n+k} (1+ H_{n+1}).\]
We have bounded from above the expected number of faults on items in $\cup_{t
\in \types(P_{i-1})} L_{t,i-1}$. We also need to add at most $h_i$ faults on
new items of types in $\types(P_{i-1})$.
\end{proof}

We summarize,
\begin{lemma} \label{lem:markd}
{\markd} is $O(\log n \max \{\log n, \log k\})$ competitive.
\end{lemma}
\begin{proof}
Each fault is counted by either $\text{charge}(P_i)$ (Proposition
\ref{prop:markd_new_ub}) or $\text{charge}(D_i)$
(Proposition~\ref{prop:markd_old_ub}) (faults on request indices in $D_i$
for items of type in $\types(P_{i-1})\setminus \types(P_i)$ are counted
twice), and by Lemma~\ref{lem:opt_lb}, we have that the expected number of
faults of {\markd} is at most
\[ \bigl (5 ( 1+H_{n+k}) + 10( 1+ (1+H_{n+1})H_{n+k}) \bigr )\cost_{\opt} .  \qquad \qed \]
\renewcommand{\qed}{}
\end{proof}

{\em Proof of Theorem~\ref{thm:lognlogk}.}
Follows immediately from Lemma~\ref{lem:marke} and Lemma~\ref{lem:markd}.
\qed

Unfortunately, the competitive ratio of a type preference
algorithm is always $\Omega(\log n \log k)$.

\begin{example} \label{examp:tp-lb}
The following example proves that the competitive ratio of a type preference
algorithm is always $\Omega(\log n \log k)$. Let $A$ be a type preference
algorithm. Let $m=n+1$ and assume there are exactly $k+1$ items from each
type. In each $P_i$ there is only one new item. At the beginning of phase
$i$, $\min D_i$ , the adversary requests all the items with the same type as
the new item, and $A$ incurs a cost of $H_{k+1}$. After that, $A$ is forced
to evict an item of a different type. The adversary chooses a type that has
the hole in it with probability at least $\frac{1}{n}$ and requests all the
items of this type each time choosing the item with maximum probability of
being a hole. This costs $A$
\[
\frac{1}{n(k+1)}+ \frac{1}{n k}+ \frac{1}{n(k-1)}+ \cdots+\frac{1}{n}=
\frac{H_{k+1}}{n}.
\]
After that, the hole is in one of $n-1$ types. Again, the adversary picks a
type that has the hole in it with probability at least $\frac{1}{n-1}$ and
requests all the items of this type each time choosing the item with maximum
probability of being a hole, which costs $A$ $H_{k+1}/(n-1)$, and so on. In
total, the expected cost for $A$ for the phase is $H_{k+1} H_n$.
\end{example}

\section{Concluding Remarks}

We have shown that the deterministic competitive ratio for $(n,k)$-companion
caching is exactly $(n+1)(k+1)-1$. We have also shown a lower bound of
$\Omega(\log n + \log k)$ and an upper bound of $O(\log n \log k)$ on the
randomized competitive ratio. We conjecture that the lower bound  we have
proven is tight. Specifically, we conjecture that the following algorithm is
$O(\log n+ \log k)$ competitive.

\paragraph*{Algorithm {\markc}:}
On a fault on item $e$ of type $t$: let $i\geq 1$ be the current
phase. If $t \notin \types(P_{i-1})$, use type eviction if
possible. Otherwise, use cache-wide eviction.

\subsection*{Acknowledgments}
This paper reflects joint work between the listed authors and Amos Fiat. 
Due to the tragic circumstances, and at Professor Fiat's insistence, 
he is not listed as an author.

\end{document}